\documentclass[12pt, letterpaper]{article}

\usepackage[T1]{fontenc}
\usepackage{amsmath, amssymb}
\usepackage{braket}
\usepackage{bm}
\usepackage{siunitx}
\usepackage{chemformula}
\usepackage[version=4]{mhchem}

\usepackage{geometry}
\usepackage{setspace}
\usepackage{graphicx}
\usepackage{float}
\newfloat{scheme}{htbp}{los}
\floatname{scheme}{Scheme}
\floatname{chart}{Chart}
\newfloat{graph}{htbp}{loh}

\usepackage{booktabs}
\usepackage{multirow}
\usepackage{array}
\usepackage{tabularx} 
\usepackage{xcolor}

\DeclareSIUnit\angstrom{\text {Å}}

\usepackage{tikz}
\usetikzlibrary{shapes.geometric, arrows.meta, positioning, calc}

\usepackage[
    backend=biber,
    style=chem-acs,
    articletitle=true,
    doi=true,
    url=true
]{biblatex}
\usepackage{hyperref}

\usepackage{authblk}

\author[1,2]{Seenivasan Hariharan}
\author[1,2]{Kareljan Schoutens}
\author[3]{Sachin Kinge}
\author[4]{Lucas Visscher}

\affil[1]{\footnotesize Institute for Theoretical Physics, University of Amsterdam, Amsterdam, The Netherlands}
\affil[2]{\footnotesize QuSoft, CWI, Amsterdam, The Netherlands}
\affil[3]{\footnotesize Toyota Motor Europe, Materials Engineering Division, Zaventem, Belgium}
\affil[4]{\footnotesize Department of Chemistry and Pharmaceutical Sciences, Vrije Universiteit, Amsterdam, The Netherlands}

\date{\footnotesize *Email: hseeni@gmail.com}

\title{Quantum Multiscale Modeling: A Hierarchy of Algorithms for Complex Chemical Systems}

\begin{document}

\maketitle

\begin{abstract}
Multiscale modeling of complex chemical systems requires algorithms that operate coherently across electronic, atomistic, mesoscopic, and continuum scales. While quantum algorithms have been proposed for each regime, no systematic framework exists to compose them across scale boundaries. Here, we identify the conditions under which fault-tolerant quantum algorithms might preserve scale-specific quantum advantages. We map quantum phase estimation, Hamiltonian simulation with Gibbs state preparation, quantum random walks, and quantum partial differential equation solvers onto electronic structure, molecular dynamics, mesoscopic kinetics, and continuum reactor physics, respectively. Crucially, these correspondences do not imply unconditional end-to-end quantum advantage; speedups depend heavily on state preparation, memory architectures, matrix conditioning, and classical readout costs. Six unresolved questions define this composition problem, illustrated via a quantum hierarchy for \ce{CO} oxidation over \ce{Pt(111)}. We propose viewing inter-scale transfer as a quantum channel composition problem at the interface of algorithm design and non-equilibrium statistical mechanics, and ask whether information loss at scale boundaries is intrinsic to multiscale modeling or merely a consequence of lossy classical transduction between algorithmic layers. The resulting roadmap suggests that multiscale quantum advantage is governed primarily by the structure of information transfer between algorithmic layers, rather than by performance at individual scales alone.
\end{abstract} 

\section{Introduction}
\label{sec:intro}

A defining challenge in computational chemistry is that the physical phenomena governing the performance of a catalyst~\cite{Bruix2019MMCatalysis}, a drug molecule~\cite{Clancy2016MMdrugs}, or a functional material~\cite{Fish2021MMmaterials} unfold simultaneously across multiple scales that differ by many orders of magnitude in both length and time.~\cite{Radhakrishnan2021surveyMM} Consider heterogeneous catalysis — the enabling chemistry behind ammonia synthesis for fertilizer production, automotive emission control, hydrogen production and utilization, carbon dioxide conversion, fuel cells, petrochemical refining, and the synthesis of sustainable fuels and commodity chemicals — as the canonical case.~\cite{Friend2017CatalysisCentral} At the electronic scale (Scale I; \SI{\sim 1}{\angstrom}, \SI{\sim 1}{\femto\second}), quantum mechanical interactions between adsorbate electrons and surface $d$-bands determine whether a molecule binds, dissociates, or passes unreacted. At the atomistic scale (Scale II; \SI{\sim 10}{\nano\meter}, \SI{\sim 1}{\nano\second}), thermal fluctuations drive surface diffusion, reconstruct the active site, and set the pre-exponential factors ($k_{0}$) that appear in Arrhenius rate expressions. At the mesoscopic scale (Scale III; \SI{\sim 1}{\micro\meter}, \SI{\sim 1}{\milli\second}), the interplay of adsorption, desorption, surface reaction, and lateral interactions governs steady-state coverages and the turnover frequency that experiments measure. At the continuum scale (Scale IV; \SI{\sim 1}{\centi\meter}, \SI{\sim 1}{\second}), species and energy transport couple with surface chemistry to determine conversion, selectivity, and operando stability.~\cite{Bruix2019MMCatalysis} No single computational method resolves all four scales in heterogeneous catalysis modeling; multiscale modeling is, fundamentally, the science of crossing the boundaries between them. While hierarchical multiscale modeling links distinct physical scales through sequential information transfer, hybrid and embedding schemes (such as quantum mechanics/molecular mechanics (QM/MM) and quantum-classical dynamics) partition a single scale, treating important regions with higher-fidelity representations. This perspective has been articulated as a general computational paradigm in multiscale modeling and simulation, where the focus lies on the systematic representation and transformation of information across scales.~\cite{Hoekstra2014MMpositionpaper} In this work, we extend this viewpoint by evaluating whether this paradigm can be formulated as a formal framework for multiscale modeling based on fault-tolerant quantum algorithms. Unlike prior multiscale quantum chemistry frameworks that focus primarily on embedding quantum resources within localized active spaces or isolated algorithmic layers~\cite{Ma2023multiscale, Capone2024VisionMM, Bickley2025extending}, our Perspective treats cross-scale information transfer itself as the central computational object.
 
The classical hierarchy for this problem is well-established. Density functional theory (DFT) supplies detailed structural information, static adsorption energies and activation barriers, while coupled-cluster methods can provide a high-accuracy refinement of the energy for small systems.~\cite{Norskov2009} To capture finite-temperature effects and structural fluctuations, ab initio molecular dynamics (AIMD) can be used to sample DFT potential energy surfaces directly by generating nuclear trajectories.~\cite{Marx2000aimd} However, these explicit vibrational effects and temperature-dependent anharmonicities are rarely transferred intact to upper scales. Instead, kinetic Monte Carlo (kMC) typically evolves the chemical reaction network (CRN) stochastically using rigid, harmonic rate constants derived from lower-scale calculations to produce coverages and turnover frequencies.~\cite{Pineda2022kinetic, Reuter2011kmc} Reactor-scale partial differential equations (PDEs) close the hierarchy by coupling surface source terms with bulk transport.~\cite{Deutschmann2015modeling} Each level is computationally demanding in its own right: electronic structure methods range from DFT, scaling formally as $O(N^3)$ and often insufficient for strongly correlated adsorbates,~\cite{Cohen2008,} to wavefunction-based approaches with steep polynomial or even exponential scaling that limits their application to small molecular clusters.~\cite{Bartlett2007cc} At the atomistic scale, molecular dynamics requires long-time sampling of nuclear trajectories to capture rare events, which routinely bounds timescales to the nanosecond or microsecond regime.~\cite{Jia2020pushing} Moving to the mesoscale, kMC simulations become highly intractable when rate-constant hierarchies span many orders of magnitude. In such cases, the simulation spends most of its time sampling rapid, reversible transitions that contribute little to the overall reaction dynamics.~\cite{Chatterjee2007kmc} Finally, macroscopic reactor-scale PDE discretizations introduce severe numerical challenges, frequently involving $10^6$–$10^8$ degrees of freedom to capture complex, multi-dimensional fluid transport and chemical species fields.~\cite{Jakobsen2008chemical} While the computational bottlenecks differ across scales, every transition necessarily involves approximating the information transferred between models.

It must be emphasized that the classical baseline is itself a moving target. Machine-learned interatomic potentials now reproduce \emph{ab initio} forces at a small fraction of the cost of explicit DFT evaluations,~\cite{Omranpour2025MLPCatal, Batatia2025foundation} effectively collapsing the cost barrier between Scales I and II, and machine-learned surrogates increasingly accelerate the construction and solution of kinetic models at Scale III.~\cite{Margraf2019systematic, Margraf2023exploring} These approaches, however, inherit the accuracy ceiling of the electronic structure data they are trained on and offer no systematic route to the strongly correlated regimes where DFT itself fails.~\cite{Kempen2026accurate} The case for quantum algorithms in the multiscale hierarchy therefore rests  not only on systematically improvable accuracy at Scale I, but equally on preserving that additional information as it propagates across successive scales.

What is rarely addressed is a structural question about the entire hierarchy: when an algorithm at one scale passes information to an algorithm at the next, what form does this information take, and should this mapping change fundamentally when both algorithms are quantum?  Quantum algorithms have been proposed to accelerate individual scales of the multiscale hierarchy,~\cite{Hariharan2024modeling} from using quantum phase estimation (QPE) to obtain high-accuracy electronic structure data,~\cite{Babbush2018lowdepth, Lee2021THC, Reiher2017elucidating} to near-optimal Hamiltonian simulation for atomistic dynamics.~\cite{Childs2021trotter, Low2017optimal} At the mesoscale, multidimensional quantum walks on mass-action system graphs provide a quantum analogue of kMC that leverages reaction network structure for Markov-chain speedups,~\cite{Jeffery2023MDQW, Hariharan2025CRN} while QSVT-based linear solvers offer exponential improvements in system size for reactor-scale PDEs.~\cite{Harrow2009, Gilyen2019QSVT, An2023} However, while each proposal is individually well-grounded, every classical algorithm in the hierarchy is merely replaced by a quantum counterpart. The critical scientific question is not whether these individual analogues exist, but how to compose them coherently across scale boundaries, shifting the objective from isolated speedups to the integration of quantum-accelerated routines into unified simulation pipelines.~\cite{Castaldo2026utility}

Even if every individual quantum subroutine achieves its intended quantum advantage, that advantage can only be realized end-to-end if the information exchanged across successive scales preserves the quantities that matter.
Rather than proposing a new quantum algorithm for a single scale, this Perspective establishes inter-scale composition as the central question in quantum multiscale modeling. Specifically, we define the scale boundary $\mathcal{I}_{k\to k+1}$ as an explicit information channel between quantum subroutines, identify six open questions governing the mathematical structure and uncertainty propagation of these channels, and map this hierarchy to a concrete case study of \ce{CO} oxidation over \ce{Pt(111)}. Ultimately, we argue that preserving end-to-end quantum advantage depends critically on the composition of these inter-scale boundaries, rather than on isolated accelerations within individual algorithmic layers. Identifying these hierarchical bottlenecks precisely, rather than focusing on modular algorithm substitutions, constitutes the central contribution of this work.

\section{The classical--quantum algorithm correspondence}
\label{sec:correspondence}

Table~\ref{tab:correspondence} presents the central idea of this Perspective. The four rows map each physical scale onto its canonical classical method and its primary quantum analogue, together with the computational complexity of the quantum analogue and, most importantly, the boundary information passed to the next coarser scale. What this mapping makes clear in the rightmost column is that the information passed upward at each boundary is, in the classical hierarchy, a lossy, classical observable—such as an energy, a rate constant, or a coverage. Our core thesis is that this classical bottleneck fundamentally can limit end-to-end quantum advantage, motivating the investigation of coherent quantum information channels where the downstream task permits.

\begin{table}[htbp]
  \centering
  \caption{%
    Fault-tolerant quantum multiscale modeling hierarchy and classical analogues. Complexities reflect the dominant bottleneck sub-task; $N$ = basis functions, $\Vert H \Vert$ = Hamiltonian norm, $\epsilon$ = target precision, and $n$ = system dimension. The information rows define the proposed scale-boundary channels that are the focus of this Perspective.\\
    (\footnotesize{Abbreviations: DFT = Density Functional Theory; CCSD(T) = Coupled-Cluster Singles and Doubles with perturbative triples; FCI = Full Configuration Interaction; AIMD = Ab initio Molecular Dynamics; kMC = kinetic Monte Carlo; CRN = Chemical Reaction Network; QPE = Quantum Phase Estimation; QRW = Quantum Random Walk; QRAM = Quantum Random Access Memory; QSVT = Quantum Singular Value Transformation; PDE = Partial Differential Equation; NVT = the canonical ensemble, keeps the number of particles (N), volume (V), and temperature (T) constant.}).}
  \label{tab:correspondence}
  \footnotesize
  \renewcommand{\arraystretch}{1.3}
  \begin{tabularx}{\linewidth}{@{} l X @{}}
    \toprule
    \textbf{Approach} & \textbf{Algorithmic framework, complexity, \& channel information} \\
    \midrule

    \multicolumn{2}{@{}l}{\textbf{SCALE I: Electronic} ($\sim$\qty{1}{\angstrom} / \qty{1}{\femto\second})} \\
    \midrule
    \textbf{Classical} & Baseline: Self-consistent field (SCF) and correlated electronic-structure calculations. \\
    \textbf{Quantum}   & QPE on qubitized Hamiltonian: $\mathcal{O}(\Vert H \Vert/\epsilon)$ queries to the block-encoding. Avoids representing of the exponentially large FCI vector; advantage conditional on preparing an initial state with non-vanishing ground-state overlap. \\
                       & \emph{Challenge:} High state-preparation overhead. \\
    \textit{Information} & Extracts ground-state adsorption energies $\Delta E_{\mathrm{ads}}$, activation barriers $E_a$, and pre-exponential factor $k_{0}$; parameterized into Arrhenius expressions for Scale II/III. \\
    
    \midrule

    \multicolumn{2}{@{}l}{\textbf{SCALE II: Atomistic} ($\sim$\qty{10}{\nano\meter} / \qty{1}{\nano\second})} \\
    \midrule
    \textbf{Classical} & Baseline: \textit{Ab initio} MD (AIMD) and canonical ensemble ($NVT$) thermal sampling. \\
    \textbf{Quantum}   & Fault-tolerant Gibbs state preparation followed by coherent Hamiltonian simulation: $\mathcal{O}(t \cdot \Vert H \Vert \cdot \mathrm{poly}\log(1/\epsilon))$ queries for real-time propagation of thermal observables; polynomial state preparation speedup. \\
                       & \emph{Challenge:} Gibbs state-preparation overhead. \\
    \textit{Information} & Extracts temperature-dependent free-energy diffusion barriers $\Delta G^\ddagger_{\mathrm{diff}}(T)$ and lateral interaction parameters $\omega_{ij}$; compiled as a coverage-dependent rate matrix. \\
    
    \midrule

    \multicolumn{2}{@{}l}{\textbf{SCALE III: Mesoscopic} ($\sim$\qty{1}{\micro\meter} / \qty{1}{\milli\second})} \\
    \midrule
    \textbf{Classical} & Baseline: Stochastic kMC and non-linear mass-action kinetics. \\
    \textbf{Quantum}   & Multidimensional QRW on mass-action system graphs via circuit theory: provable query speedups via QRAM to decide species reachability, sample reachable states, and approximate steady-state fluxes under network perturbations. \\
                       & \emph{Challenge:} Dependence on QRAM and CRN topology sensitivity. \\
    \textit{Information} & Extracts steady-state surface coverages $\theta_i^*$, turnover frequencies (TOF), and selectivities; injected as local source terms $\dot{\omega}_i$ into Scale IV. \\
    
    \midrule

    \multicolumn{2}{@{}l}{\textbf{SCALE IV: Continuum} ($\sim$\qty{1}{\centi\meter} / \qty{1}{\second})} \\
    \midrule
    \textbf{Classical} & Baseline: Computational Fluid Dynamics (CFD) and Macroscopic Reactor PDEs. \\
    \textbf{Quantum}   & QSVT-based sparse linear PDE solvers and quantum ODE/PDE integrators: $\mathcal{O}(\mathrm{poly}\log(n) \cdot \kappa \cdot \log(1/\epsilon))$ gate complexity for accessing global observables of suitably structured sparse systems, compared with $\mathcal{O}(n)$ classical scaling. \\
                       & \emph{Challenge:} Speedup conditional on condition number $\kappa$ and efficient observable readout. \\
    \textit{Information} & Resolves macroscale reactor concentration profiles $c_i(\mathbf{x},t)$, temperature fields $T(\mathbf{x},t)$, and integrated terminal yields/selectivities. \\

    \bottomrule
  \end{tabularx}
\end{table}

In conventional multiscale workflows, the output of one computational scale is typically converted into classical observables before being passed to the next scale. This measurement-mediated interface is entirely appropriate when the downstream model requires only those observables. However, measurement can also discard information contained in the quantum representation of the intermediate calculation, including amplitudes, phases, correlations, or entanglement. When the downstream computation can operate directly on this representation, a quantum-native composition of adjacent scales may avoid the intermediate conversion

$$
\text{quantum state}\rightarrow\text{classical data}\rightarrow\text{new quantum state}.
$$

The potential benefit of such coherent composition is therefore not simply the elimination of state-preparation cost. Rather, it is the possibility of preserving and processing information across computational scales without first reducing the intermediate result to a prescribed set of classical observables. Whether this provides a computational advantage is problem dependent and requires comparison with the costs of measurement, classical processing or reconstruction, and subsequent state preparation. Thus, intermediate measurement should not be regarded as intrinsically disadvantageous; instead, the appropriate interface depends on which information is required by the subsequent scale and on the relative cost of classical and coherent composition.

\subsection{Classical computational scaling across the multiscale hierarchy}

\begin{table*}[htbp]
\centering
\caption{\textbf{Classical computational scaling across the four scales.}
The classical cost at each scale is determined by both the cost of an elementary computational step and the number of steps, events, samples, or solver iterations required to achieve the desired physical accuracy. The expressions shown represent characteristic cost models rather than universal asymptotic complexities. Here, MP2 denotes second-order M\o ller--Plesset perturbation theory, $N$ denotes the system size, $N_{\rm traj}$ the number of independent trajectories, $N_{\rm steps}$ the number of integration steps, $R$ the number of reaction channels, $S$ the number of kinetic states or species, $N_{\rm events}$ the number of reaction events, $N_{\rm t}$ the number of time steps, $N_{\rm iter}$ the number of solver iterations, $N_{\rm DOF}$ the number of discretized degrees of freedom and $C_{(\cdot)}$ the cost of the corresponding elementary operation.}

\label{tab:classical_scaling}
\renewcommand{\arraystretch}{1.35}
\begin{tabular}{p{0.32\textwidth} p{0.65\textwidth}}
\hline
\textbf{Scale and classical calculation} &
\textbf{Representative classical cost} \\
\hline

\textbf{I: Electronic}
\newline
SCF and correlated electronic-structure calculations
&
\begin{tabular}[t]{@{}l@{}}
$C_{\rm ES}(N)$; e.g.,
\\
$O(N^3)$ for conventional pure DFT,
\\
$O(N^4)$ for hybrid DFT,
\\
$O(N^5)$ for MP2,
\\
$O(N^7)$ for CCSD(T)
\end{tabular}
\\[0.8em]

\textbf{II: Atomistic}
\newline
Ab initio MD (AIMD) and canonical ensemble (NVT) thermal sampling.
&
\begin{tabular}[t]{@{}l@{}}
$N_{\rm traj}N_{\rm steps}C_{\rm force}(N)$, with
\\
$C_{\rm force}(N)=O(N^3)$ for DFT-based AIMD; 
\\
for classical force fields, 
\\
$O(N^2)$ for naive pairwise interactions, 
\\
approaching $O(N)$ for short-range interactions and
\\
approximately $O(N\log N)$ for particle-mesh Ewald methods
\end{tabular}
\\[0.8em]

\textbf{III: Mesoscopic}
\newline
Stochastic kMC and non-linear mass-action kinetics.
&
\begin{tabular}[t]{@{}l@{}}
$N_{\rm events}C_{\rm event}(R)$ for kMC,
\\
or
\\
$N_t C_{\rm RHS}(S,R)N_{\rm iter}$
\\
for time-dependent kinetic equations
\end{tabular}
\\[0.8em]

\textbf{IV: Continuum}
\newline
Computational Fluid Dynamics (CFD) and Macroscopic Reactor PDEs.
&
\begin{tabular}[t]{@{}l@{}}
$N_{\rm DOF}N_{\rm iter}C_{\rm operator}$,
\\
where the scaling depends on the
\\
discretization and numerical solver
\end{tabular}
\\

\hline
\end{tabular}
\end{table*}

The classical computational cost of the four scales is not determined solely by the formal scaling of an individual computational operation. At each scale, the total cost depends on both the cost of an elementary computational step and the number of steps, events, statistically independent samples, or solver iterations required to achieve the target physical accuracy. We therefore characterize the classical
baseline in terms of these two contributions, as summarized in Table~\ref{tab:classical_scaling}.

The resulting expressions should be interpreted as representative cost models rather than universal asymptotic complexities. At Scale~I, formal polynomial scaling provides a useful characterization of the dominant electronic-structure cost. For example, conventional pure and hybrid density-functional calculations exhibit characteristic cubic and quartic scaling, respectively, while correlated methods
such as (multireference) perturbation theory and coupled cluster approaches exhibit substantially steeper scaling with system size~\cite{Jensen2017introduction}. In practice, these formal scalings are however often substantially reduced using density fitting or Cholesky decomposition~\cite{Pedersen2024Cholesky}, local correlation methods~\cite{Nagy2024LocalCorrelation} and tensor compression techniques~\cite{Glaser2024TensorDecomposition}, which make the actual scaling less transparent and to some extent system-specific. At the subsequent scales, the picture is still less clear. The computational bottleneck then depends increasingly on the number of trajectories, events, samples, or numerical iterations required to obtain a converged physical observable. Consequently, the nominal cost of an individual operation does not by itself characterize the total computational burden.

For Scale~II, for example, the cost of molecular dynamics can be written schematically as
\begin{equation}
C_{\rm MD}
\sim
N_{\rm traj}N_{\rm steps}C_{\rm force}(N).
\label{eq:md_classical_cost}
\end{equation}
The force-evaluation cost may scale as $O(N^2)$ for naive pairwise interactions, approach $O(N)$ for short-range interactions with appropriate neighbor-list methods, or approximately $O(N\log N)$ for particle-mesh treatments of long-range electrostatics~\cite{George2022review} (for AIMD, $C_{\rm force}$ is instead dominated by the $O(N^3)$ cost of each DFT evaluation).  However, this per-step scaling does not capture the full cost of obtaining statistically converged thermodynamic or free-energy quantities. The number of statistically independent configurations is reduced by temporal correlations, with the statistical inefficiency determined by the integrated autocorrelation time. Consequently, for a target statistical error $\epsilon$, the total sampling cost can be expressed schematically as
\begin{equation}
C_{\rm MD}
=
O\!\left[
N_{\rm traj}
\frac{\tau_{\rm int}}{\Delta t}
\epsilon^{-2}
C_{\rm force}(N)
\right],
\label{eq:md_sampling_cost}
\end{equation}
where $\tau_{\rm int}$ is the integrated autocorrelation time and $\Delta t$ is the integration timestep. Thus, although force evaluation may have relatively low polynomial scaling in $N$, the computational cost of converging a free-energy or rare-event observable can instead be dominated by the long-time sampling required to decorrelate configurations and adequately explore the relevant configuration space.

At Scale~III, the classical cost is associated with the propagation of a kinetic network over the relevant physical time and with the resolution of the reaction channels that contribute to the dynamics. For kinetic Monte Carlo (kMC), a representative cost is
\begin{equation}
C_{\rm kMC}
\sim
N_{\rm events}C_{\rm event}(R),
\label{eq:kmc_classical_cost}
\end{equation}
where $N_{\rm events}$ is the number of reaction events required to reach the target physical time~\cite{Lin2019scaling} and $C_{\rm event}(R)$ denotes the cost of constructing or sampling from the set of $R$ reaction channels. Importantly, $N_{\rm events}$ is not determined solely by the number of reactions in the network: it also depends on the reaction propensities and on the separation between fast and slow processes. Large disparities in reaction timescales can therefore lead to a large number of microscopic events being required to resolve a much slower chemical process ~\cite{Pineda2022kinetic}. For deterministic kinetic descriptions, the corresponding cost can be expressed schematically as
\begin{equation}
C_{\rm ODE}
\sim
N_t\,C_{\rm RHS}(S,R)\,N_{\rm iter},
\label{eq:ode_classical_cost}
\end{equation}
where $N_t$ is the number of integration steps, $S$ is the number of kinetic states or species, and $N_{\rm iter}$ accounts for the iterations required by the numerical integration scheme. Thus, for Scale~III, network size, kinetic stiffness, and the physical timescale that must be resolved can all contribute to the classical
computational cost.

At Scale~IV, the classical cost is determined primarily by the discretization of the reactor equations and the numerical solution of the resulting linear or nonlinear system. A representative cost model is
\begin{equation}
C_{\rm reactor}
\sim
N_{\rm DOF}N_{\rm iter}C_{\rm operator},
\label{eq:reactor_classical_cost}
\end{equation}
where $N_{\rm DOF}$ is the number of spatial and, where applicable, temporal degrees of freedom, $N_{\rm iter}$ is the number of iterations required for convergence, and $C_{\rm operator}$ is the cost of applying the discretized differential operator or solving the associated linear system. The resulting scaling depends strongly on
the spatial discretization, boundary conditions, stiffness of the coupled equations, and choice of numerical solver~\cite{Raimondeau2002recent, Saad2003iterative}. Direct sparse solvers, Krylov methods, and multigrid methods can therefore exhibit substantially different scaling with $N_{\rm DOF}$. Consequently, a single universal polynomial complexity is not representative of the continuum scale; the dominant classical cost is instead determined by the size and conditioning of the discretized problem together with the convergence requirements of the solver.

Taken together, these scaling considerations highlight that the classical cost of multiscale modelling is governed by different computational bottlenecks at different levels of the hierarchy. Electronic-structure calculations are often dominated by the scaling of the underlying many-electron method, whereas atomistic simulations can be limited by long-time sampling, mesoscopic kinetics by the number of reaction events and separation of timescales, and continuum models by the size and conditioning of the discretized equations. This distinction is important when assessing potential quantum advantage: an improvement in the asymptotic scaling of an individual subroutine does not necessarily translate into an end-to-end multiscale advantage unless the dominant sampling, state-preparation, conditioning, communication, and measurement costs are also addressed.

\subsection{A quantum multiscale hierarchy for CO oxidation over Pt(111)}
\label{sec:casestudy}
To illustrate how the proposed hierarchy appears in practice, we consider catalytic \ce{CO} oxidation over a close-packed platinum surface (Pt(111)),

\begin{equation}
\ce{2CO(g) + O2(g) ->[Pt(111)] 2CO2(g)}
\label{eq:overall}
\end{equation}
which proceeds predominantly through a Langmuir--Hinshelwood mechanism involving adsorption, oxygen dissociation, surface diffusion, and the subsequent reaction of adsorbed intermediates. In its minimal representation, the elementary reaction network consists of
\begin{align} 
\mathrm{CO(g) + *} &\rightleftharpoons \mathrm{CO^*}, \\ 
\mathrm{O_2(g) + 2*} &\rightleftharpoons \mathrm{2O^*}, \\ 
\mathrm{CO^* + O^*} &\rightarrow \mathrm{CO_2(g) + 2*}. 
\end{align}
\noindent where \(*\) denotes an empty surface site. This reaction represents a prototypical benchmark in heterogeneous catalysis whose mechanism, surface intermediates, and non-equilibrium kinetic behavior have been studied extensively through classical electronic-structure calculations, kinetic modeling, and operando experiments.~\cite{Reuter2011kmc,Bruix2019MMCatalysis} 

This minimal network provides a concrete example of how observables extracted at one scale become inputs to subsequent scales. Furthermore, this workflow could be scaled up to handle the exact structural complexities, such as single-atom or magnetic active sites, identified by recent industry-academic roadmaps for industrial catalyst screening.~\cite{Hariharan2024modeling} In particular, we can formalize the information transfer across the Scale I--IV hierarchy as the systematic compilation of a multiscale parameter vector:

\begin{equation}
\mathcal{\lambda} = \begin{Bmatrix}
\Delta E_{\mathrm{ads}}, & E_a, & \Delta G^\ddagger_{\mathrm{diff}}(T) , & \omega_{ij}, & k_j, & \theta_i^*, & \mathrm{TOF}
\end{Bmatrix}
\end{equation}

where these components correspond respectively to adsorption energies, activation barriers, diffusion barriers, lateral interaction parameters, rate constants, steady-state surface coverages, and turnover frequencies. The explicit purpose of this case study is to show precisely how physical information crosses each scale boundary, what specific chemical observables are transferred, and where the structural composition questions of quantum multiscale modeling manifest (Section~\nameref{sec:questions}).

\paragraph{Scale I — Electronic structure of \ce{CO/Pt(111)}.}
The initial task is determining the ground-state adsorption energies, activation barriers, and transition-state geometries for elementary events including \ce{CO} adsorption, \ce{O2} dissociation, and \ce{CO2} formation. Classically, these quantities are obtained from density functional theory (DFT) or correlated wavefunction methods. The drawback is well-documented: approximate exchange-correlation functionals introduce severe localized errors on transition-metal surfaces (e.g., the infamous DFT puzzle of incorrect \ce{CO} site-preference on \ce{Pt(111)}),~\cite{Cohen2008,Feibelman2001} while systematically improvable classical methods scale exponentially.

In the fault-tolerant hierarchy, QPE provides a systematically improvable alternative for high-accuracy electronic structure, evaluating eigenvalues with a query complexity scaling as $\mathcal{O}(\Vert H \Vert/\epsilon)$ and potentially avoiding explicit classical representation of exponentially large wavefunction vectors.~\cite{Babbush2018lowdepth,Lee2021THC} To formalize this interface mathematically for the \ce{CO/Pt(111)} system, we consider a position-dependent formulation where the electronic Hamiltonian is block-conditional on a nuclear configuration register~\cite{Kassal2008polynomial, Ward2009preparation}:

\begin{equation}
\hat{H}_{\mathrm{elec}} = \sum_i |i\rangle\langle i| \otimes \hat{H}_i
\end{equation}

\noindent where $|i\rangle$ indexes a discrete grid of nuclear configurations corresponding to local adsorption geometries on the platinum slab, and $\hat{H}_i$ represents the corresponding electronic Hamiltonian at that geometry. This expression provides an idealized coherent description of a discretized Born–Oppenheimer energy landscape, in which the nuclear degrees of freedom are represented quantum mechanically in a discretized configuration basis \(\{|i\rangle\}\), while the electronic Hamiltonian is parameterized by each nuclear configuration. By preparing a spatial superposition over these surface configurations and executing conditional QPE to produce a coherent electronic energy register~\cite{Obrien2019multiple}, the algorithm outputs an entangled state vector across the joint nuclear, electronic, and energy registers:

\begin{equation}
|\psi_{\mathrm{I}}\rangle = \sum_i c_i |i\rangle |\phi_i\rangle | \Delta E_{\mathrm{ads}}^{(i)} \rangle
\label{eq:coefficients}
\end{equation}

\noindent where $|\phi_i\rangle$ represents the electronic state and the energy register stores the precise adsorption energy eigenvalues $\Delta E_{\mathrm{ads}}^{(i)} $ corresponding to each specific configuration $|i\rangle$. Thus, Scale I does not assume a classical sampling of fixed molecular geometries; the nuclear configuration register remains in coherent superposition until the Scale I$\rightarrow$II transformation.

In principle, the coherent energy register produced by QPE could serve directly as an input to quantum sampling procedures rather than being measured immediately. Rather than collapsing this state projectively via intermediate classical measurement—which can reintroduce severe input/output bottlenecks—this entangled register is instead used coherently as the control register for a Scale~II Gibbs reweighting procedure $U_{\mathrm{Gibbs}}$~\cite{Chowdhury2016quantum}. By appending an ancilla qubit and executing a coherent conditional rotation controlled on the energy register, 

\begin{equation}
    |i\rangle |\phi_i\rangle |\Delta E_{\mathrm{ads}}^{(i)}\rangle |0\rangle \mapsto |i\rangle |\phi_i\rangle |\Delta E_{\mathrm{ads}}^{(i)}\rangle
    \left( e^{-\beta \Delta \widetilde{E}^{(i)}/2}|0\rangle
    + \sqrt{1 - e^{-\beta \Delta \widetilde{E}^{(i)}}}\,|1\rangle \right)
\end{equation}

\noindent with $\Delta \widetilde{E}^{(i)} = \Delta E_{\mathrm{ads}}^{(i)} - E_{\min}$ shifted by the minimal register energy to keep the rotation well-defined. Post-selecting the ancilla on $|0\rangle$ and tracing out the electronic and energy registers yields the mixed density operator over the nuclear configuration space:

\begin{equation}
\rho_{\mathrm{II}} = \frac{1}{\widetilde{Z}} \sum_i |c_i|^2\, e^{-\beta \Delta \widetilde{E}^{(i)}} |i\rangle\langle i| = \frac{1}{\widetilde{Z}'} \sum_i |c_i|^2\, e^{-\beta \Delta E_{\mathrm{ads}}^{(i)}} |i\rangle\langle i|,
\label{eq:density-operator}
\end{equation}

\noindent where $\widetilde{Z} = \sum_i |c_i|^2 e^{-\beta \Delta \widetilde{E}^{(i)}}$, $\widetilde{Z}' = \sum_i |c_i|^2 e^{-\beta \Delta E_{\mathrm{ads}}^{(i)}}$, and $\beta = (k_B T)^{-1}$. The constant energy shift $E_{\min}$ factors out and cancels exactly under normalization, leaving the relative thermal weights invariant. The Scale~I amplitudes $c_i$ do not disappear: $\rho_{\mathrm{II}}$ reduces to the exact Gibbs state only when the input superposition is uniform ($|c_i|^2 = 1/N$); otherwise the thermal weights remain modulated by the upstream preparation. Moreover, the post-selection succeeds with probability $P_{\mathrm{succ}} = \sum_i |c_i|^2 e^{-\beta \Delta \widetilde{E}^{(i)}}$, which, in unfavorable cases, can become exponentially small in $\beta \cdot \mathrm{spread}(\Delta E_{\mathrm{ads}})$, requiring $\mathcal{O}(1/\sqrt{P_{\mathrm{succ}}})$ rounds of amplitude amplification. This construction therefore bypasses intermediate classical measurement and re-encoding of the energy information, but at a quantifiable boundary cost—precisely the query-complexity trade-off that Question~Q1 identifies as the central object of the coherent channel $\mathcal{I}_{\mathrm{I}\rightarrow \mathrm{II}}$.

\paragraph{Scale II — Adsorbate dynamics on \ce{Pt(111)}.}
At the atomistic scale, the objective shifts from isolated ground-state energies to computing the temperature-dependent parameters that govern the mobility and spatial distribution of species across the \ce{Pt(111)} surface. Specifically, the model requires the free-energy diffusion barriers $\Delta G^\ddagger_{\mathrm{diff}}(T)$ for \ce{CO^*} hopping between adjacent top and bridge sites, and the lateral interaction parameters $\omega_{ij}$ that quantify how neighboring \ce{CO^*} or \ce{O^*} spectator species alter local adsorption and reaction energetics.

Rather than running trillions of classical force evaluations to log classical trajectories, the fault-tolerant quantum framework uses the density matrix passed from Scale~I, $\rho_{\mathrm{II}}$, as the initial quantum representation from which a Gibbs state associated with the interacting multi-adsorbate Hamiltonian is prepared via block-encoding methods. By configuring this state to represent co-adsorbed \ce{CO^*} and \ce{O^*} at varying coordinate distances, the quantum computer deploys quantum algorithms for free-energy estimation to evaluate coverage-dependent free-energy differences and interaction energies. Coherent, real-time Hamiltonian simulation is subsequently used to sample the finite-temperature dynamics of the interacting adsorbate system.

From this simulation pipeline, the target observables are extracted via quantum expectation values, yielding the temperature-dependent free-energy diffusion barriers $\Delta G^\ddagger_{\mathrm{diff}}(T)$ and a matrix of lateral interaction coefficients $\omega_{ij}$. This directly raises the first Question~Q1: if the structural relaxation of the underlying \ce{Pt(111)} lattice happens on a timescale comparable to the adsorbate diffusion hops, the coarse-graining step injects memory effects, requiring $\mathcal{I}_{\mathrm{II} \to \mathrm{III}}$ to be formalized as a non-Markovian quantum channel rather than a memoryless rate assignment.

\paragraph{Scale III — \ce{CO} oxidation kinetics.}
The classical mesoscopic description accounts for spatial heterogeneity and correlations across the catalyst surface by evolving the master equation over the combinatorial state space of the \ce{Pt(111)} lattice network. Here, the rate matrix $\mathbf{W}$ is compiled using the parameters passed from the previous scale, where local event frequencies are updated dynamically based on coverage: 

\begin{equation}
    k_{\alpha} = \frac{k_{B}T}{h} \exp[-\beta\,\Delta G^\ddagger_{\alpha}(T, \boldsymbol{\theta})]
\end{equation}

\noindent with the total coverage-dependent activation free energy formalized as 

\begin{equation}
    \Delta G^\ddagger_{\alpha}(T, \boldsymbol{\theta}) = \Delta G^\ddagger_{\alpha,0}(T) + \sum_l\omega_{\alpha l}\theta_l .
\end{equation}

\noindent Here, $\Delta G^\ddagger_{\alpha,0}(T)$ is the intrinsic reference free-energy barrier of elementary event $\alpha$, and $\omega_{\alpha l}$ denotes the energetic barrier correction arising from lateral interactions with neighboring species $l$. For catalytic \ce{CO} oxidation, extreme mathematical stiffness arises because the rate constant for rapid \ce{CO^*} diffusion hops is several orders of magnitude larger than the rate constant for the rare, slow recombination step to form \ce{CO2}. This causes classical kinetic Monte Carlo (kMC) to expend a dominant fraction of computational effort simulating rapid, non-reactive diffusion back-and-forth steps, contributing substantially to the computational cost through high-frequency reversible events without significantly advancing the overall reaction coordinate.

In the quantum multiscale hierarchy, this stiffness bottleneck is addressed by representing the chemical reaction network (CRN) as a multidimensional quantum random walk (QRW) mapped directly onto the mass-action system graph—rather than the full lattice configuration graph—using circuit theory equivalents and QRAM.~\cite{Hariharan2025CRN} The walk operates on species-count states of the mass-action network, and the corresponding electrical network mappings can accelerate the computation of steady-state properties under favorable graph and spectral conditions. An important qualification applies here: the mass-action representation is a mean-field description in species numbers, whereas lattice kMC resolves explicit spatial correlations---including the coverage-dependent lateral interaction parameters from Scale~II, which enter the kinetic model through the barrier corrections $\omega_{\alpha l}$. Recovering spatial resolution within the quantum walk framework requires either enlarging the network with site-resolved pseudo-species or developing new walk constructions on lattice configuration graphs. Quantifying what is lost in the mean-field contraction, and at what cost it can be restored, is part of the composition problem formalized below in Question~Q5.

Rather than executing a full classical readout of the state space---or converging to a stationary distribution as a classical Markov chain would---the quantum walk encodes steady-state properties in the spectral structure of its walk operator for suitably constructed reversible or graph-embedded Markov processes, from which quantum phase and amplitude estimation extract two specific macro-kinetic observables: the steady-state surface coverages $\theta_{\ce{CO}}^*$ and $\theta_{\ce{O}}^*$, and the net turnover frequency ($\text{TOF}$) of emitted $\ce{CO2}$ per active site. As highlighted in Question~{Q5}, the core structural challenge is verifying that the sparse, anisotropic graph topology of the actual $\ce{CO}/\ce{O}/\ce{Pt}(111)$ surface network preserves the required QRW spectral gap when bound to these specific input and output parameters. 

\paragraph{Scale IV — The \ce{CO} oxidation reactor.}
At the macroscopic continuum scale, the catalytic surface is embedded into a physical chemical reactor model (e.g., a stagnation-flow boundary layer reactor or a packed-bed channel over the Pt catalyst). Here, gas-phase transport and mass transfer limitations are coupled to the local surface kinetics via conservation partial differential equations (PDEs):
\begin{equation}
\frac{\partial c_i}{\partial t} + \nabla\cdot \mathbf{J}_i = \dot{\omega}_i(c_{\ce{CO}}, c_{\ce{O2}}, T, \boldsymbol{\theta}^*)
\end{equation}
where $c_i$ represents the macroscopic gas-phase concentrations of \ce{CO}, \ce{O2}, and \ce{CO2}. The coupling between the microscopic catalytic scales and the reactor model is introduced through the volumetric chemical source terms $\dot{\omega}_i = \nu_i \Gamma_s a_s \mathrm{TOF}$, where $\nu_i$ is the stoichiometric coefficient, $\Gamma_s$ is the molar density of active sites per unit catalyst area, $a_s$ represents the catalyst surface-area density, and $\mathrm{TOF}$ is the quantum-estimated surface turnover frequency obtained within the reduced kinetic representation of Scale~III. 

Under a hybrid classical–quantum Newton-QSVT approach, spatial discretization of the reactor geometry on a mesh of $n$ nodes reduces each linearized implicit step of the nonlinear fluid dynamics solver to a sparse linear system $A\mathbf{x} = \mathbf{b}$ solved via the Quantum Singular Value Transformation (QSVT).~\cite{Gilyen2019QSVT,An2023} The true quantum advantage at this scale is shaped by Question~Q6: if the engineering objective requires the full spatial visualization of the concentration field $c_i(\mathbf{x},t)$ throughout the entire reactor volume, the classical readout bottleneck can eliminate the practical advantage associated with the favorable polylogarithmic system-size dependence of the idealized QSVT linear solver. 

Instead, the pipeline preserves its potential quantum advantage by avoiding full tomography, using quantum expectation measurements to extract a single global macroscopic engineering metric: the total integrated \ce{CO2} production rate at the reactor outlet, defined by the boundary operator:
\begin{equation}
\dot{N}_{\ce{CO2, out}} = \int_{\text{outlet}} \mathbf{J}_{\ce{CO2}} \cdot d\mathbf{A}
\end{equation}
By compressing the multiscale quantum simulation down to a single target scalar observable, the global pipeline completes its handoff while avoiding classical reconstruction of the high-dimensional intermediate representations generated at the electronic, atomistic, and mesoscopic scales.

\begin{figure}[htbp]
\centering
\includegraphics[width=\textwidth]{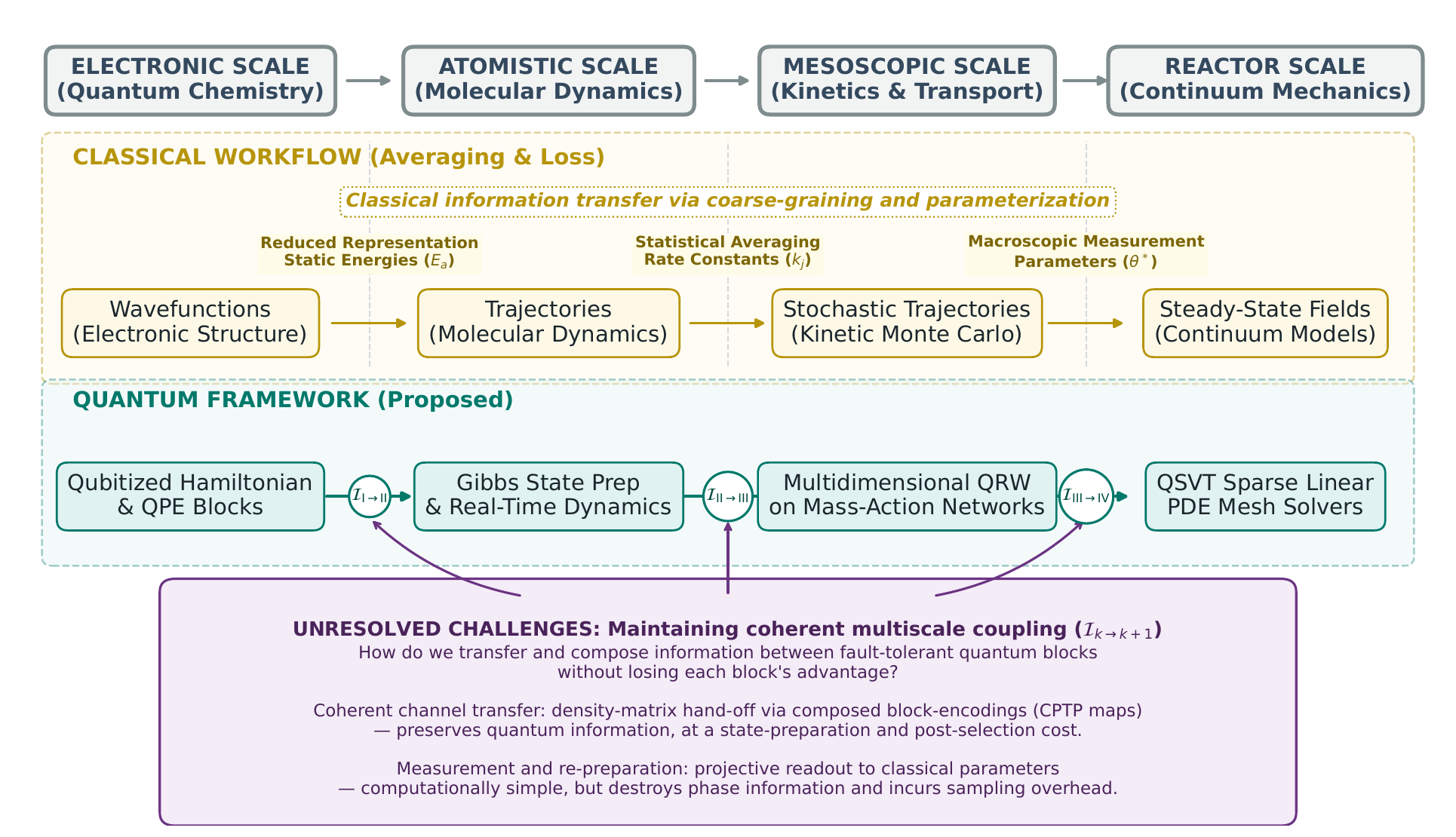}
\caption{
Comparison of classical and fault-tolerant quantum workflows across electronic, atomistic, mesoscopic, and continuum scales. Classical approaches progressively compress information through averaging, reduced representations, and measurement. The proposed quantum hierarchy aims to compose coherent transformations across scales using fault-tolerant quantum algorithms. Interface nodes $\mathcal{I}_{k\to k+1}$ denote unresolved challenges in maintaining coherent multiscale coupling.
}
\label{fig:multiscale_quantum_pipeline}
\end{figure}

The \ce{CO} oxidation on \ce{Pt(111)} case study reinforces that the central difficulty lies not in proposing individual quantum counterparts, but in composing them: resolving how cross-scale channels ($\mathcal{I}_{k \to k+1}$) handle information transfer without destroying the individual advantage of each computational layer. This architectural divergence---and the structural questions it raises at each interface---is mapped conceptually in Fig.~\ref{fig:multiscale_quantum_pipeline}.

\section{Generalizing the hierarchy across chemical systems}

The \ce{CO} oxidation pipeline is one instance of a construction that applies to any chemical system whose physics spans the four scales of Table~\ref{tab:correspondence}. In this section we abstract away from \ce{Pt(111)} and state, for each scale, the general computational task, the fault-tolerant algorithm that addresses it, and the complexity and hardware conditions under which its localized advantage survives. The emphasis shifts from \emph{what} information crosses each boundary, which is established above, to \emph{at what cost} it does so for arbitrary reaction networks and reactor geometries.

\paragraph{Scale I — Electronic structure.}
At the electronic scale (Scale I), the central computational task is solving the many-electron Schr\"{o}dinger equation for an active site embedded in a larger model such as a lattice with periodic boundary conditions. Density Functional Theory (DFT) handles this at polynomial cost (formally $\mathcal{O}(N^3)$), while wavefunction-based coupled-cluster theory restricts calculations to clusters of tens of atoms due to its steep $\mathcal{O}(N^7)$ scaling, and exact Full Configuration Interaction (FCI) scales exponentially with system size. Among these approaches, DFT remains the undisputed workhorse of heterogeneous catalysis modeling. 

As illustrated for \ce{CO/Pt(111)} above, approximate exchange-correlation functionals fail precisely for the strongly correlated, spin-polarized surfaces that catalyze industrially important reactions.~\cite{Cohen2008, Hariharan2024modeling} To break these classical limitations, fault-tolerant quantum algorithms have been proposed at this scale, with QPE positioned to compute precise electronic ground states and activation barriers. The query complexity of QPE on a qubitized Hamiltonian scales as $\mathcal{O}(\Vert H \Vert/\epsilon)$, where $\Vert H \Vert$ is the spectral norm and $\epsilon$ the target precision.~\cite{Babbush2018lowdepth, Lee2021THC} But this does not constitute an unconditional exponential speedup. The total cost is governed by the overlap of the prepared initial state with the true ground state, and whether this overlap decays with system size for chemically relevant problems remains actively debated.~\cite{Lee2023evidence} A landmark resource estimate by Reiher et al.\ showed that QPE could determine the catalytic mechanism of \ce{N2} fixation by the FeMo-cofactor of the nitrogenase enzyme, a prototypically strongly correlated active site, using on the order of $10^{2}$ logical qubits and $\sim 10^{15}$ T gates. Although daunting, these estimates define the regime at which quantum advantage becomes chemically relevant.~\cite{Reiher2017elucidating} Subsequent algorithmic developments have reduced these requirements by several orders of magnitude; for example, spectrum-amplification estimates for FeMoco require  $\sim$1.1--1.5$\times10^3$ logical qubits and $\sim3\times10^8$--$10^9$ Toffoli gates (Table \ref{tab:multiscale_resources}), making fault-tolerant quantum simulations of chemically relevant systems increasingly  promising.~\cite{Goings2022CYPP450, Low2025SOSSA}

The information transmitted through the Scale~I $\rightarrow$ Scale~II
interface---ground-state adsorption energies $\Delta E_{\mathrm{ads}}$ andactivation barriers $E_a$---is conventionally condensed into classical Arrhenius- or Eyring-type rate expressions. In a fully fault-tolerant multiscale paradigm, the structural question becomes whether this information must be first extracted via classical measurement or can be utilized directly in Scale II.

The dissipative and coherent implementations of this interface were made explicit in the case study: projective measurement of the entangled register of Eq. \ref{eq:coefficients} collapses it to classical scalars, whereas the conditional Gibbs-state construction of Eq. \ref{eq:density-operator} transmits it coherently at a quantifiable post-selection cost.

\paragraph{Scale II — Molecular dynamics.}
At the atomistic scale, the focus shifts from static ground-state energies to the real-time evolution of nuclear positions and thermal sampling: how do adsorbates diffuse, how does the surface reconstruct under reaction conditions, and what are the coverage-dependent lateral interactions? Classical \textit{ab initio} molecular dynamics (AIMD) resolves this but is strictly limited to picosecond-to-nanosecond trajectories on systems of hundreds of atoms by the $\mathcal{O}(N^3)$ cost of each classical density functional theory (DFT) force evaluation. 

To capture the finite-temperature behavior that defines true catalyst performance, a zero-temperature ($T=0$) ground-state Hamiltonian is insufficient. Instead, the fault-tolerant analogue requires explicit quantum Gibbs state preparation ($\rho \propto e^{-\beta H}$) to establish thermal equilibrium within the canonical ensemble (NVT). Once the thermal mixed state is prepared via block-encodings, coherent real-time Hamiltonian simulation samples and evolves observables associated with this thermal ensemble. Qubitization and linear-combination-of-unitaries (LCU) methods achieve a near-optimal query complexity of $\mathcal{O}(t \cdot \Vert H \Vert \cdot \mathrm{poly}\log(1/\epsilon))$, yielding a near-linear time dependence and exponential precision tracking.\cite{Childs2012LCU} 

Concretely, these fault-tolerant algorithms bridge the gap to higher scales by translating abstract wavefunctions into physical observables that govern long-timescale kinetics. First, rather than relying on classical functionals, the quantum computer evaluates analytical atomic forces $\mathbf{F}_I = -\nabla_{\mathbf{R}_I} E(\mathbf{R})$ via quantum gradient estimation algorithms with polynomial scaling. Second, by preparing Gibbs states across varying reaction coordinates, the framework utilizes thermodynamic integration to directly compute Helmholtz free energy barriers:

$$\Delta A = -\frac{1}{\beta} \ln\left(\frac{Z_1}{Z_0}\right)$$

Third, real-time Hamiltonian simulation allows the direct extraction of dynamical correlation functions, such as the velocity autocorrelation function $\langle \mathbf{v}(t) \cdot \mathbf{v}(0) \rangle = \mathrm{Tr}\left( \rho \, e^{iHt/\hbar} \mathbf{v} e^{-iHt/\hbar} \mathbf{v} \right)$, yielding transport coefficients whose accuracy is limited only by the simulated Hamiltonian and the estimation precision, rather than by force-field or mean-field approximations.

The resulting temperature-dependent free-energy diffusion barriers $\Delta G^\ddagger_{\mathrm{diff}}(T)$ (obtained from free-energy differences $\Delta A$; for adsorbed phases the $p\Delta V$ distinction between Helmholtz and Gibbs barriers is negligible) and lateral interactions $\omega_{ij}$ (mapped via cluster expansions of co-adsorbed thermal states) leave Scale II to compile the coverage-dependent rate matrix of Scale III. Here, local hopping rates are parameterized via Transition State Theory:

\[ k_{i \to j} = \frac{k_B T}{h} \exp\left( -\frac{\Delta A^\ddagger(\omega_{ij})}{k_B T} \right) \]

If the underlying Born--Oppenheimer surface is itself evaluated via QPE, the full Scale I/II pipeline becomes a coherent quantum computation. Yet even in this best case, the boundary itself remains classical: the exponentially complex correlations sampled within Scale II are compressed into a handful of scalar rates $k_{i \to j}$ before crossing to Scale III. Whether this compression is a physical necessity of coarse-graining or an artifact of the classical kMC target and what a coherent alternative would cost, is precisely the trade-off formalized in Q1.

\paragraph{Scale III — Mesoscopic kinetics.}
The mesoscopic kinetic baseline evolves the master equation of a chemical reaction network (CRN):
\begin{equation}
  \frac{d\mathbf{P}}{dt} = \mathbf{W}\mathbf{P},
  \label{eq:master}
\end{equation}
where $\mathbf{P}(t)$ is the probability vector over all microstates of the surface and $\mathbf{W}$ is the transition rate matrix whose off-diagonal elements are proportional to the rate constants $k_j$ derived from Scale I/II.\cite{Pineda2022kinetic, Reuter2011kmc} For a surface with $N_{\mathrm{sites}}$ sites, the state space grows exponentially, making direct classical integration intractable. As shown in the \ce{CO} oxidation example, classical kMC faces severe bottlenecks when rate constants span many orders of magnitude.

Alternative quantum approaches to solving Eq.~\eqref{eq:master} generally rely on mapping the stochastic master equation to a continuous-variable system via Schrödingerization or quantum linear systems algorithms (QLSA)~\cite{Jin2023quantum, Berry2017linear}, which require expanding non-Hermitian operators into higher-dimensional dilated Hamiltonians. While these approaches offer polynomial advantages for deterministic differential integration, they struggle with the intrinsically non-linear structures generated by macroscopic mass-action constraints. 

To overcome this, Scale III is replaced by a multidimensional quantum random walk (QRW) mapped directly onto the non-linear mass-action system graph using circuit-theoretic equivalents.\cite{Hariharan2025CRN} Rather than relying on generic, unstructured graph-walk speedups which fail to track steady-state distributions, this framework maps the CRN onto multidimensional electrical networks utilizing alternative Kirchhoff's and Ohm's laws via Quantum Random Access Memory (QRAM).\cite{Hariharan2025CRN} 

It is critical to acknowledge the hardware overhead this architecture imposes. The proposed query speedups rely strictly on the availability of QRAM for efficient species reachability queries and state sampling. Fault-tolerant QRAM architectures pose physical and engineering challenges. If QRAM cannot be realized with overheads that scale sub-linearly with the CRN state space, the kinetic sampling bottleneck will remain unbroken. This highlights a critical hardware-software co-design constraint: mesoscopic quantum speedups are as dependent on memory-access architecture as they are on algorithmic query complexity. 

The general CRN topology is encoded by the stoichiometric matrix $\mathbf{U} = [\nu'_{ij} - \nu_{ij}]_{N \times M} $ across the reaction pathways:
\begin{equation}
  \sum_i \nu_{ij} S_i \;\xrightarrow{k_j}\; \sum_i \nu'_{ij} S_i, \quad j = 1, \ldots, M,
  \label{eq:CRN}
\end{equation}
This specific multidimensional architecture yields provable quantum query speedups to decide species reachability, sample highly probable reachable states under network perturbations, and approximate steady-state fluxes over classical kMC.\cite{Hariharan2025CRN} 

To transition this mesoscopic picture to the macroscopic continuum description of Scale IV, the steady-state surface coverages $\theta_i^*$ and reaction fluxes $J_j^*$ are extracted from the final quantum state preparation. The net reaction rate $R_j$ for each pathway is computed directly from these quantum-accelerated steady-state fluxes. These rates are then contracted with the stoichiometric matrix to yield the local volumetric net production rates:

\begin{equation}
  \dot{\omega}_i = \sum_{j=1}^{M} (\nu'_{ij} - \nu_{ij}) R_j
\end{equation}

These scalar source terms $\dot{\omega}_i$ successfully compress the discrete, exponentially scaled microstate probabilities $\mathbf{P}(t)$ of the surface down to local, deterministic chemical source profiles. By passing $\dot{\omega}_i$ forward, Scale III provides the mathematical boundary conditions required to couple the microscopic quantum kinetics directly into the fluid dynamics and transport equations of the macroscopic reactor scale.

\paragraph{Scale IV — Reactor scale.} 
At the macroscale, species mass and energy conservation yield a system of coupled, non-linear macroscopic partial differential equations (PDEs):
\begin{equation}
  \frac{\partial c_i}{\partial t} + \nabla \cdot \mathbf{J}_i = \dot{\omega}_i(\{ \theta_j^* \}, T, \{c_k\}),
  \label{eq:reactor}
\end{equation}
where $c_i$ is the gas-phase concentration of species $i$, $\mathbf{J}_i$ is its spatial transport flux, and $\dot{\omega}_i$ is the volumetric net production rate explicitly compiled and passed forward from the Scale III chemical reaction network. 

When discretized on a spatial mesh of $n$ nodes, the full reactor model becomes a highly non-linear algebraic system. To port this to a quantum architecture, two distinct paths exist. The first is a hybrid framework: linearizing the non-linear source terms about the current operating point via a classical outer Newton loop, reducing the inner problem to a large, sparse linear system of size $n \times n$. Quantum Singular Value Transformation (QSVT) sparse linear system solvers map this inner task into the fault-tolerant domain,~\cite{Gilyen2019QSVT} executing the matrix inversion $A\mathbf{x} = \mathbf{b}$ with a gate complexity of $\mathcal{O}(\mathrm{poly}\log(n) \cdot \kappa \cdot \log(1/\epsilon))$, where $\kappa$ is the condition number. Alternatively, fully quantum non-linear PDE algorithms utilize Carleman linearization or non-linear Schrödingerization to map the non-linearities directly into an infinite-dimensional linear Bose-Einstein or dilated state space.~\cite{An2023, Krovi2023improvedquantum} Carleman-based approaches, however, carry a strict validity condition: efficient truncation of the linearized hierarchy is guaranteed only when dissipation dominates the non-linearity,~\cite{Liu2021carleman} a condition that strongly non-linear Arrhenius source terms will generally violate and
that must be verified for any given reactor regime. Both pathways can yield an exponential reduction in scaling with respect to the spatial mesh size $n$ compared to classical direct solvers. Quantum ODE/PDE integrators expand this advantage cleanly to time-dependent propagation.~\cite{Berry2014} 

Crucially, this algorithmic advantage remains strictly conditional on three major bottlenecks. First, for the hybrid Newton-QSVT approach, the classical computer must re-compile and re-load the block-encoding of the Jacobian matrix $A$ at every iteration, presenting a severe data-loading overhead. Second, the speedup assumes the condition number $\kappa$ remains well-behaved and does not scale exponentially with the mesh resolution $n$ or fluid turbulence. Third, and most fundamentally, the entire framework is bound by the quantum input/output (I/O) problem. While the quantum solver efficiently prepares a state $\ket{x}$ proportional to the spatial solution vector, extracting the full macroscopic concentration field $c_i(\mathbf{x},t)$ via quantum state tomography requires $\mathcal{O}(n)$ measurements, which completely nullifies the exponential algorithmic speedup.

True quantum advantage at Scale IV is therefore strictly governed by a hardware-software-engineering co-design principle: the target engineering objective must not require full field visualization. Instead, it must target a global scalar property—such as integrated terminal reactor yield, total reactant conversion, or global entropy production—that can be extracted efficiently via quantum expectation value estimation ($\langle x | \hat{O} | x \rangle$) without collapsing the full spatial state vector.

\subsection{State of the art and resource requirements}
The current state of the art is considerably more mature for the individual quantum algorithmic components than for their integration into a complete multiscale workflow. Resource estimates have been reported for electronic-structure calculations and related quantum simulation tasks, whereas fault-tolerant resource requirements for chemically realistic finite-temperature dynamics, reaction-network kinetics, and continuum-scale calculations remain less developed. Moreover, existing estimates generally treat the inputs from preceding scales as externally supplied. Thus, an end-to-end resource estimate for a chemically consistent quantum multiscale workflow, including the transfer of information between scales, has not yet been established. Table \ref{tab:multiscale_resources} summarizes representative demonstrations, reported resource estimates, and the remaining requirements at each scale.

\begin{table}[htbp]
\centering
\caption{\textbf{Representative fault-tolerant resource estimates across
quantum multiscale modelling.} Reported resources are
problem- and algorithm-dependent.}
\label{tab:multiscale_resources}
\small
\setlength{\tabcolsep}{4pt}
\renewcommand{\arraystretch}{1.22}
\begin{tabularx}{\columnwidth}{@{}l X@{}}
\toprule
\textbf{Scale} & \textbf{Representative example and FT resource status} \\
\midrule

\textbf{I. Electronic}
&
\textbf{FeMoco ground-state energy} \cite{Low2025SOSSA}:
$1.1$--$1.5\times10^{3}$ logical qubits and
$10^{8}$--$10^{9}$ Toffoli gates, with
$\sim3.2$--$4.5\times10^{6}$ physical qubits reported for
the corresponding surface-code estimates.
\\[2pt]

\textbf{II. Atomistic}
&
\textbf{Nuclear motion (30D)} \cite{Szczepanik2026nuclear}:
$<300$ logical qubits and $\sim3$ months on a hypothetical
$1$\,MHz FT processor. FT-MD resources additionally scale with
trajectory length, sampling and repeated force evaluations.
\\[2pt]

\textbf{III. Mesoscopic}
&
\textbf{Reaction networks} \cite{Hariharan2025CRN, Kabengele2026kinetics}:
4--5 qubit demonstrations. No end-to-end FT estimate is
currently available for a chemically realistic CRN; resources
depend on network size, oracle cost, mixing/sampling and precision.
\\[2pt]

\textbf{IV. Continuum}
&
\textbf{Advection--diffusion} \cite{Ingelmann2024advection,Novikau2025advection}:
14-qubit demonstration; $10^{6}$ grid points require at least
20 data qubits for index encoding. Total FT resources depend on
sparsity, condition number, block encoding and precision.
\\[2pt]

\midrule

\textbf{Integrated}
&
\textbf{Full multiscale workflow}:
$C_{\mathrm{FT}}^{\mathrm{total}}
\simeq \sum_{i=\mathrm{I}}^{\mathrm{IV}} N_i C_i$,
where $C_i$ is the cost per quantum subroutine and $N_i$ its
number of invocations. The current $\sim10^{3}$-logical-qubit,
$10^{8}$--$10^{9}$-Toffoli estimates therefore represent the
cost of an electronic-structure call rather than the complete
multiscale workflow.
\\

\bottomrule
\end{tabularx}
\end{table}

A quantitative estimate of the hardware requirements for this complete multiscale workflow is necessarily uncertain because the resources depend on algorithmic choices, target precision, state preparation, data access, and the transfer of information between scales. Table \ref{tab:logical_resources} therefore provides indicative logical-qubit requirements for each scale and for the combined CO/Pt(111) workflow, based on currently available resource estimates and algorithmic scaling.

The individual estimates should be regarded as order-of-magnitude indicators rather than definitive requirements. In particular, the resources of the four scales cannot simply be added, since calculations at one scale may need to be repeated as inputs from another scale are updated. The overall requirement is therefore strongly dependent on the frequency of inter-scale communication and the extent to which intermediate quantities can be compressed. For the complete workflow, we estimate an indicative requirement of approximately $10^3$–$10^6$ logical qubits. This range should be viewed as a projection based on the current state of the art rather than a prediction of future hardware requirements.

\begin{table*}[h]
\centering
\caption{\textbf{Indicative logical-resource requirements across the four scales of the quantum multiscale workflow.}
The resource ranges are representative rather than universal and are intended to indicate the order of magnitude of logical-qubit requirements under fault-tolerant assumptions. Actual requirements depend strongly on target precision, Hamiltonian simulation method, state preparation, oracle and data-access costs, and the structure of the multiscale coupling.}
\label{tab:logical_resources}
\renewcommand{\arraystretch}{1.35}
\begin{tabular}{p{0.2\textwidth} p{0.2\textwidth} p{0.53\textwidth}}
\hline
\textbf{Scale}
& \textbf{Indicative logical resources}
& \textbf{Main resource drivers}
\\
\hline
Electronic & $10^{2}$--$10^{4}$
& Number of orbitals, target precision, Hamiltonian simulation,
ansatz or state-preparation requirements, and phase-estimation
precision \\

Atomistic & $10^{2}$--$10^{5}$
& System size, simulation time, force evaluation, quantum state
preparation, sampling requirements, and accuracy of dynamical
observables
\\[0.6em]

Mesoscopic & $10^{3}$--$10^{5}$
& Network size, number of reaction channels, oracle or data-access
requirements, state preparation, and target precision
\\[0.6em]

Continuum & $10^{3}$--$10^{6}$
& Discretization size, conditioning of the discretized equations,
linear-system or differential-equation solver requirements,
state preparation, and readout
\\[0.6em]

\textbf{Multiscale workflow} & $10^{3}$--$10^{6}$
& Inter-scale coupling, repeated calls between scales, precision
requirements, data transfer, state preparation, and the extent to
which intermediate quantities can be compressed
\\

\hline
\end{tabular}

\end{table*}

Although \ce{CO} oxidation on \ce{Pt(111)} provides a concrete illustrative example, the proposed hierarchy is applicable to a broader range of catalytic and electrochemical systems, including ammonia synthesis, C--H activation, \ce{CO2} conversion, hydrogen and oxygen evolution and reduction reactions, and catalysis on transition-metal alloys, single-atom catalysts, and magnetic surfaces. Related multiscale structures also arise in molecular-interaction problems relevant to drug discovery, where electronic structure, molecular dynamics, free-energy sampling, and kinetic models can be coupled. The specific quantum algorithms and inter-scale interfaces will depend on the chemistry and its dominant computational bottlenecks; the framework therefore provides a strategy for identifying where quantum
subroutines could be incorporated, rather than implying quantum advantage across all scales or application domains.

\section{Open questions in quantum multiscale modeling}
\label{sec:questions}
Having established that a rigorous quantum analogue exists within each distinct scale, the overarching scientific problem shifts from individual algorithmic isolation to systemic architectural coupling.
Table~\ref{tab:questions} distills this challenge into six interconnected open questions. Q1 and Q2 concern the boundaries themselves---the mathematical structure of the inter-scale channel and whether all boundaries in the hierarchy are equally difficult to implement; Q3 evaluates whether a unified algebraic framework can compose the quantum subroutines across these boundaries; Q4 addresses how uncertainty propagates through them; and Q5 and Q6 confront the end-to-end question: the total query complexity of the composed pipeline and the structural conditions under which it retains a net quantum advantage. Rather than treating these interfaces as mere data-passing steps, we state each problem
with mathematical precision as a necessary first step toward formalizing a true calculus of quantum multiscale simulation.

More formally, we define the inter-scale interface $\mathcal{I}_{k\to k+1}$ as a transformation acting on a quantum state (or density operator) $\rho_{k}$, associated uncertainty budget $\epsilon_{k}$, and admissible observable set $\mathcal{O}_k$, producing inputs for the downstream algorithm. Classical multiscale methods correspond to projective measurement followed by parameter compression, whereas coherent interfaces preserve quantum information through CPTP maps or generalized quantum instruments.

\begin{table}[htbp]
  \centering
  \caption{%
    Six open questions defining the quantum multiscale composition problem. $\mathcal{I}_{k\to k+1}$ denotes the inter-scale quantum or classical information channel from Scale $k$ to Scale $k+1$.
  }
  \label{tab:questions}
  \renewcommand{\arraystretch}{1.5}
  \begin{tabular}{p{0.05\linewidth} p{0.89\linewidth}}
    \toprule

    \textbf{Q1} &
    \textbf{Mathematical structure of the inter-scale channel.} \newline
    What is the minimal mathematical description of $\mathcal{I}_{k\to k+1}$: a memoryless CPTP map, a generalized quantum instrument, or a non-Markovian channel carrying temporal correlations injected by cross-scale coarse-graining? When adjacent scales both execute fault-tolerant quantum algorithms, must the channel terminate in a projective measurement that collapses the state to classical parameters, or can quantum information be transmitted coherently via density-matrix mappings---and what is the query complexity cost of each paradigm? \\

    \textbf{Q2} &
    \textbf{Heterogeneity of inter-scale boundaries.} \newline
    Are all channels $\mathcal{I}_{k\to k+1}$ equally difficult to implement? The lower boundaries transfer microscopic, Hamiltonian-level information between algorithms sharing a common physical representation, whereas the upper boundaries exchange averaged statistical and continuum-level quantities; the central boundary must bridge these two regimes. Does this representational divide render some channels naturally amenable to coherent composition while others remain irreducibly dissipative---and do hybrid quantum--classical pipelines inherit the same asymmetry at their classical--quantum boundaries? \\

    \textbf{Q3} &
    \textbf{QSVT as a unifying computational primitive.} \newline
    Can the QSVT serve as a single unifying primitive to implement all core quantum subroutines---QPE, Hamiltonian simulation, quantum walks, and linear system inversions? If so, does this algebraic unification reduce scale-bridging to the clean composition of block-encoded operators within hybrid or fully coherent architectures? \\

    \textbf{Q4} &
    \textbf{Uncertainty propagation across quantum channels.} \newline
    Quantum amplitude estimation offers a generic quadratic speedup for uncertainty quantification (UQ) across individual scales. However, does the inter-scale quantum channel $\mathcal{I}_{k\to k+1}$ inherently amplify, preserve, or dampen irreducible quantum measurement uncertainty under cross-scale coarse-graining?  \\

    \textbf{Q5} &
    \textbf{End-to-end query complexity of the global pipeline.} \newline
    What is the total query complexity of the integrated four-scale quantum pipeline as a function of target global precision $\varepsilon$, global system size $N$, and total simulation time $T$? Is this multi-layer complexity provably sub-polynomial in the corresponding classical multiscale baseline, and which scale boundaries dictate the dominant bottleneck? \\

    \textbf{Q6} &
    \textbf{Conditions for end-to-end quantum advantage.} \newline
    If physical or informational constraints force $\mathcal{I}_{k\to k+1}$ to be implemented via classical measurement and re-preparation, what structural conditions on the individual subroutines guarantee that the composed pipeline retains a net quantum advantage? Under what conditions does the sampling overhead of classical readout at the boundary destroy the speedup achieved within individual layers? \\

    \bottomrule
  \end{tabular}
\end{table}

Questions Q1–Q6 should be interpreted as open theoretical problems rather than established properties of quantum multiscale architectures. Wherever examples are provided below, they serve as illustrative constructions rather than proofs of sufficiency or necessity.

\paragraph{Q1 — Mathematical structure of the inter-scale channel.}
The physical act of cross-scale coarse-graining discards microscopic degrees of freedom, transforming a closed, unitary quantum system at Scale $k$ into an open quantum system at the boundary interface. This transition requires a precise description of the resulting informational channel $\mathcal{I}_{k \to k+1}$. Classical multiscale protocols universally rely on a memoryless Markovian assumption: information is compressed into static scalar parameters (e.g., an energy or a rate constant) under the premise that fast environmental correlations relax instantaneously. In the fault-tolerant quantum computing domain, however, it remains open whether a memoryless CPTP map or a generalized quantum instrument is mathematically sufficient to represent this interface. If the quantum correlations generated within strongly correlated adsorbate systems decay on a timescale comparable to or longer than the
downstream coarse-grained dynamics, the interface must instead be modeled as a non-Markovian quantum channel possessing temporal memory. Defining the boundary between Markovian decay and structured quantum memory is the foundational step toward a rigorous scale-crossing calculus.

Whatever its mathematical form, the channel must also be assigned a physical implementation, and here the classical template for multiscale modeling is unambiguous: every classical scale boundary realizes $\mathcal{I}_{k \to k+1}$ as a lossy, dissipative transduction, in which a projective measurement extracts a classical scalar that is injected manually into the next layer. While measurement is computationally free in standard workflows, this design can introduce bottlenecks on a quantum architecture. If Scale $k$
produces a multi-particle state $|\psi\rangle$ encoding a probability
distribution over surface configurations, projective measurement of
configuration probabilities irreversibly destroys phase information that cannot be reconstructed at Scale $k+1$ without re-running Scale $k$, while re-preparing an equivalent state downstream may, depending on the problem, incur a state-preparation cost that offsets the localized quantum speedup. The alternative, transmitting the density matrix directly to Scale $k+1$ as a coherent input, demands that the downstream algorithm be explicitly engineered to accept unmeasured quantum inputs. Tracking the query complexity tradeoff between these two paradigms, state-collapsing measurement versus coherent channel transmission, is the second half of the structural question that Q1 poses.

\paragraph{Q2 — Heterogeneity of inter-scale boundaries.}
The four scale boundaries of the hierarchy are not equal. At the lower boundary ($\mathcal{I}_{\mathrm{I} \to \mathrm{II}}$), both algorithms operate on microscopic, Hamiltonian-level representations of the same physical system: eigenstates, forces, and thermal density operators share a common mathematical language, which is precisely why a coherent construction such as the conditional Gibbs-state preparation of the \ce{CO/Pt(111)} case study is conceivable at all. At the upper boundary ($\mathcal{I}_{\mathrm{III} \to \mathrm{IV}}$), both sides already have averaged, statistical quantities, such as  coverages, fluxes, and source terms, whose transfer as classical scalars sacrifices little, since the coarse-graining has largely been paid for upstream. The central boundary ($\mathcal{I}_{\mathrm{II} \to \mathrm{III}}$) is qualitatively different: it must convert microscopic dynamical information into the statistical parameterization of a master equation, crossing the representational divide between mechanics and kinetics. We conjecture that this central boundary is where coherent transfer is simultaneously most valuable, where the exponentially
large configuration space has not yet been compressed, and most difficult to engineer. The open question is whether this asymmetry is intrinsic to the representational divide or merely an artifact of the currently available algorithmic toolbox, and whether hybrid quantum--classical pipelines, in which some layers remain classical, inherit the same asymmetry at their classical--quantum boundaries.

\paragraph{Q3 — QSVT as a unifying computational primitive.}
The Quantum Singular Value Transformation (QSVT) establishes that the seemingly disparate algorithms defining our multiscale tiers---QPE, Hamiltonian simulation, quantum walks, and linear system solvers---are structurally unified under the algebraic umbrella of polynomial transformations of block-encoded matrices.\cite{Gilyen2019QSVT, Martyn2021grand} Given a standard block-encoding of a matrix $A$ within a unitary $U_A$ such that $(\langle 0| \otimes I) U_A (|0\rangle \otimes I) = A/\alpha$, QSVT can execute any chosen polynomial transformation $p$ on the singular values of $A$ using only $\mathcal{O}(\text{deg}(p))$ calls to $U_A$. Within this context, QPE represents phase estimation on a target block-encoding; Hamiltonian simulation applies a Chebyshev approximation of $e^{-iHt}$; the mesoscopic walk executes a polynomial that amplifies the graph's target subspace; and the inner reactor PDE step evaluates a polynomial that inverts the singular values of the sparse linear subsystem. 

If these core algorithmic blocks can be compiled as specialized QSVT polynomials over respective block-encoded operators, the scale-bridging problem mathematically simplifies to the arithmetic composition of block-encodings.\cite{Gilyen2019QSVT} However, the massive structural and algebraic disparity between a highly correlated molecular electronic Hamiltonian (Scale I) and a sparse continuum transport matrix (Scale IV)—alongside the requirement for an outer classical Newton loop at the reactor scale—means a single, monolithic end-to-end quantum circuit is structurally infeasible. Instead, a viable unified QSVT architecture requires a sequence of distinct, scale-specific block-encodings interconnected by coherent quantum routing and hybrid classical data-loading subroutines.

\paragraph{Q4 — Uncertainty propagation across quantum channels.} 
In classical multiscale modeling, parameter uncertainties—such as functional-induced errors in DFT activation barriers—propagate nonlinearly through the hierarchy, driving exponential variance in predicted macroscopic rates. While quantum amplitude estimation provides a generic quadratic speedup for uncertainty quantification (UQ) within any individual layer, reducing the classical sampling overhead from $\mathcal{O}(\epsilon^{-2})$ to $\mathcal{O}(\epsilon^{-1})$, composing these bounds across scales poses a fundamentally different mathematical challenge. The critical open question is how irreducible quantum measurement uncertainty propagates when transmitted through the coarse-grained inter-scale boundary $\mathcal{I}_{k\to k+1}$, and whether this open quantum channel inherently amplifies, preserves, or dampens errors compared to classical non-linear propagation.

More generally, if inter-scale channels satisfy $\epsilon_{k+1} = f_k(\epsilon_k)$, where $f_k$ denotes the uncertainty transformation induced by the inter-scale channel $\mathcal{I}_{k \to k+1}$, then quantum multiscale uncertainty propagation reduces to characterizing the composed map:
\begin{equation}
  \epsilon_{\mathrm{global}} = f_n \circ f_{n-1} \circ \cdots \circ f_1 (\epsilon_0).
\end{equation}
The mathematical properties of the transformations $f_{k}$ therefore determine whether uncertainty is amplified, preserved, or damped under successive coarse-graining across scales.

\paragraph{Q5 — End-to-end query complexity of the global pipeline.}
The individual quantum speedups distributed across the multiscale layers span highly diverse functional dependencies: polynomial scaling in basis functions ($\mathcal{O}(\Vert H \Vert/\epsilon)$), real-time propagation bounds ($\mathcal{O}(t \cdot \Vert H \Vert)$), and polylogarithmic matrix inversion scales ($\mathcal{O}(\mathrm{poly}\log(n) \cdot \kappa)$). In classical engineering, the dominant computational bottleneck resides at the mesoscopic step (Scale III), where the exponential state space of the master equation forces stochastic sampling via kMC. 

A closed-form bound on the total query complexity of the composed four-scale pipeline does not yet exist, and deriving one is non-trivial: the scaling parameters that dominate each layer, such as $\|H\|$, spectral gap $\Delta$, and condition number $\kappa$, are not independent, and their joint dependence on global system size $N$ and target precision $\varepsilon$ remains unknown. Establishing even a partial bound—for example, at the Scale I/II interface alone—would constitute meaningful progress toward a full pipeline complexity theorem.

Resolving this overarching complexity bound fundamentally requires characterizing the spectral properties of the specific chemical reaction network (CRN) at Scale III, as these govern how the localized quantum walk speedups behave when coupled to adjacent scales. Whether the multidimensional quantum walk framework on mass-action system graphs can decisively break the mesoscopic bottleneck under chemically realistic, sparse topologies with deeply heterogeneous rate hierarchies remains to be demonstrated.\cite{Hariharan2025CRN} Evaluating the total query complexity of the integrated pipeline requires solving a dual theoretical and algorithmic problem: deriving the strict graph properties and spectral gaps governing mass-action chemical networks under steady-state perturbations, and proving that the compiled quantum walk preserves its query advantages when bound to the data inputs and outputs of its adjacent scales.

\paragraph{Q6 — Conditions for end-to-end quantum advantage.}
Beyond the aggregate complexity of Q5, the end-to-end advantage of the composed pipeline remains highly sensitive to boundary degradation. If physical or informational constraints force an inter-scale interface $\mathcal{I}_{k \to k+1}$ to remain a classical intermediary, error propagation cascades nonlinearly. For example, if a Scale I QPE routine requires $\text{Time}_{\mathrm{I}}$ gate operations to isolate an activation barrier to precision $\epsilon_{\mathrm{I}}$, this value is classically processed via a highly non-linear, ill-conditioned Arrhenius expression ($k_j \propto e^{-E_a/k_B T}$) before parameterizing the Scale III quantum walk.\cite{Pineda2022kinetic, Reuter2011kmc} 

Small variations in $\epsilon_{\mathrm{I}}$ propagate exponentially through the pipeline, meaning the localized benefit of a low $\text{Time}_{\mathrm{I}}$ only accrues if downstream sensitivities are tightly bounded. Composing multiple quantum layers under a global target precision constraint $\varepsilon = \sum_k \epsilon_k$ thus transforms into a non-trivial, multi-variable error allocation problem. To prevent order-of-magnitude downstream rate distortions at typical catalytic reaction temperatures ($T \sim 500\text{ K}$), this error budgeting requires constraining the Scale~I algorithmic precision to chemical accuracy ($\epsilon_{\mathrm{I}} \leq 1.6~\mathrm{mHa} \approx 0.043~\mathrm{eV} \approx k_B T$ at $T = 500~\mathrm{K}$). This formalizes a rigorous mathematical boundary where the maximum permissible global pipeline error $\varepsilon$ dictates strict upper bounds on the acceptable logical qubit overhead and runtime thresholds for the underlying electronic oracles.

Whether coherent inter-scale coupling provides an accuracy advantage over measurement followed by re-preparation remains an open question. Formal frameworks for composing quantum subroutines, including on superpositions of inputs~\cite{Jeffery2022quantumsubroutinecomposition, Jeffery2025composingquantumalgorithms}, and quantum algorithms coupling microscopic and macroscopic reaction–diffusion dynamics ~\cite{Lockwood2025multiscale} provide a basis for addressing this question. To date, there is no chemically realistic numerical demonstration showing that retaining coherence across scale boundaries improves the accuracy of a chemically relevant observable. The potential advantage would arise if measurement discards correlations or other information that is relevant to downstream calculations; if the measured quantities form a sufficient representation of the information required at the next scale, measurement and re-preparation may be equally accurate. Establishing when coherent coupling provides a measurable benefit, and whether that benefit justifies its additional resource requirements, is therefore an important direction for future research.

\section{Outlook}
\label{sec:outlook}

The framework developed in this Perspective identifies four fault-tolerant quantum algorithms --- quantum phase estimation, coherent Hamiltonian simulation with Gibbs state preparation, multidimensional quantum walks, and QSVT-based sparse linear solvers --- as the natural mathematical counterparts of the classical multiscale hierarchy for complex chemical systems including heterogeneous catalysis. We have formalized six open questions whose resolution will determine whether these algorithms can be systematically composed into a unified pipeline that preserves individual localized advantages to deliver a net quantum speedup over state-of-the-art classical multiscale implementations. While this framework is explicitly fault-tolerant in scope, near-term hybrid approaches — such as variational quantum eigensolvers at Scale I or quantum-classical surrogate models at Scale III — may serve as stepping stones that stress-test the inter-scale composition interfaces identified here before fault-tolerant hardware matures.

The central mathematical challenge, crystallized by Questions Q1 and Q2, is the rigorous characterization of the inter-scale information channel $\mathcal{I}_{k \to k+1}$ when adjacent layers operate fully within the quantum domain. The algebraic machinery of QSVT provides a highly promising blueprint for this challenge; by defining an arithmetic for combining block-encoded operators, it offers a path to express inter-scale boundaries as composed block-encodings rather than informationally destructive classical measurements. Concurrently, formal frameworks for the composition of quantum algorithms and nested subroutines, such as those developed by Jeffery provide the foundational query complexity machinery required to analyze how errors and oracle calls propagate across nested quantum structures.\cite{Jeffery2025composingquantumalgorithms, Jeffery2022quantumsubroutinecomposition} Whether these algorithmic tools can yield a universal composition theorem, establishing the necessary and sufficient conditions under which cross-scale quantum information transfer remains lossless with respect to end-to-end query complexity, remains the foundational open problem this Perspective defines.

A critical dimension of this composition problem is \textit{uncertainty quantification} (UQ), which must be evaluated not as an isolated problem, but as an intrinsic feature of open quantum channels. In classical multiscale modeling, uncertainties in DFT activation barriers (typically $\pm$\,\qty{0.1}{\electronvolt}) propagate nonlinearly through Arrhenius expressions, resulting in exponential sensitivity in downstream kMC rate constants and predicted turnover frequencies. Quantum amplitude estimation offers a generic quadratic speedup for UQ across every individual scale, reducing the classical sampling cost from $\mathcal{O}(\epsilon^{-2})$ to $\mathcal{O}(\epsilon^{-1})$.\cite{Brassard2002} This suggests a future analogue of classical sensitivity analysis in which quantum multiscale pipelines are optimized by allocating precision budgets $\epsilon_{k}$ across scales under fixed logical-resource constraints.
However, the deeper question highlighted by our framework (Question Q4) is how irreducible quantum measurement uncertainty propagates across a boundary that is itself an open quantum channel. Determining whether the channel $\mathcal{I}_{k \to k+1}$ inherently amplifies, preserves, or dampens uncertainty under cross-scale coarse-graining represents a vital mathematical constraint that directly shapes the answers to Questions Q1, Q2, and Q6.

Two additional technical qualifications on the pipeline deserve explicit statement. First, the multidimensional quantum walk at Scale III operates specifically on the non-linear mass-action system graph of the CRN via circuit theory analogies.\cite{Hariharan2025CRN} The multi-dimensional product structure of this state space is highly non-generic; it possesses explicit algebraic features that distinguish it from unstructured random walks on arbitrary graphs. Consequently, evaluating the global query complexity in Question Q5 requires exploiting this specific topological structure alongside QRAM data access architectures. 

Second, the QSVT-based solver at Scale IV applies strictly to the \textit{linearized} reactor PDE system arising at each Newton iteration of a global non-linear solver. Because the full macroscopic continuum reactor model is non-linear, the quantum advantage claimed at Scale IV is strictly conditional: it assumes the global classical non-linear solver spends its dominant computational cost in the repeated solution of large, sparse linear subsystems. This condition generally holds for ultra-fine, well-resolved reactor meshes, but must be verified case-by-case.

Throughout this Perspective, all quoted quantum advantages are conditional rather than universal: preserving asymptotic improvements requires assumptions regarding state preparation, memory access, spectral properties, conditioning, and observable extraction that must ultimately be validated for chemically realistic systems. The value of the framework presented here is to explicitly define the fault-tolerant research agenda for \emph{quantum multiscale modeling}. The core hurdle facing quantum multiscale modeling is not demonstrating that a quantum analogues exist at each scale; they do, though improvements in those single-scale algorithms are both needed and expected. The true challenge is demonstrating that the interfaces between those scales can be engineered to carry information without causing an informational collapse that completely destroys the quantum advantage individual layers offer. Posing that question with mathematical precision is the first necessary step toward its resolution.

\subsection*{Data availability}
No primary research data were generated or analyzed in this Perspective; all results discussed are available in the cited literature. \\

\subsection*{Author contributions}
S.H. conceived the idea of quantum multiscale modeling and initiated the project (Conceptualization) and prepared the initial draft of the manuscript (Writing – Original Draft). K.S., S.K., and L.V. contributed to the development and refinement of the manuscript through critical discussion, revision, and co-writing (Writing – Review \& Editing). K.S. obtained and managed the project funding (Funding Acquisition, Project Administration). S.K. provided project funding from Toyota Motor Europe and contributed to project administration (Funding Acquisition, Project Administration). L.V. provided supervision and contributed to project administration (Supervision, Project Administration). All authors discussed the concepts presented in the manuscript and approved the final version.

\subsection*{Conflict of interest}
The authors declare no conflict of interest. 


\printbibliography

\end{document}